\documentstyle[12pt,psfig]{article}
\newcommand{\ba}{\begin{eqnarray}}
\newcommand{\ea}{\end{eqnarray}}
\newcommand{\be}{\begin{equation}}
\newcommand{\ee}{\end{equation}}

\begin{document}
\begin{titlepage}
\begin{flushright}

hep-ph/9902368
 {\hskip.5cm}\\
\end{flushright}
\begin{centering}
\vspace{.3in}
{\bf  String Unification at Intermediate Energies:}\\
{\bf Phenomenological Viability and Implications}\\
\vspace{2 cm}
{G.K. Leontaris$^1$ and N.D. Tracas$^2$} \\ \vskip 1cm

{\it {$^1$Physics Department, University of Ioannina\\ Ioannina,
GR45110, GREECE}}\\
{\it {$^2$Physics Department, National Technical
University \\ 157 73 Zografou, Athens, GREECE}}\\

\vspace{1.5cm} {\bf Abstract}\\
\end{centering}
\vspace{.1in}
Motivated by the fact that the string scale can be many orders
of magnitude lower than
the Planck mass, we investigate the required  modifications in the MSSM
$\beta$--functions in order to achieve intermediate ($10^{10-13}$GeV)
scale unification, keeping the traditional logarithmic running of the
gauge couplings. We  present examples of string unified models with
the required extra matter for such a unification while we also check
whether other MSSM properties (such as radiative symmetry breaking)
are still applicable.

\vspace{2cm}
\begin{flushleft}
February 1999
\end{flushleft}
\hrule width 6.7cm \vskip.1mm{\small \small}
 \end{titlepage}
\section{Introduction}

Recent developments in string theory have revealed the interesting
possibility that the string scale $M_{string}$ may be much lower
than the Planck mass $M_{P}$.  According to a suggestion \cite{w}
the string scale  could be identified  with the minimal
unification scenario scale $M_{string}\sim 10^{16}$GeV. It was
further noted that, if extra dimensions remain at low
energies\cite{i,Ly,a,i1,i2,i3}, unification of gauge couplings may
occur at scales as low as a few TeV \cite{a}. However, it is not
trivial to reconcile this scenario with all the low energy
constraints\cite{i4,add,ny}. 
 Recently \cite{li,imr} it was further proposed that
in the weakly--coupled Type I string vacua the string scale can
naturally  lie in some intermediate energy, $10^{10-13}$GeV, which
happens to be the geometrical mean of  the $M_{P}$ and weak,
$M_W$, scales (i.e. $M_{string} \sim \sqrt{M_W M_{P}}\sim 10^{11}$GeV).
It is a rather interesting fact that the possibility of
intermediate scale unification  was also  shown to appear in the
context of Type IIB theories\cite{AP}. This scenario has the
advantage that this intermediate scale does not need the
power--like running of the gauge couplings in order to achieve
unification. Appearance of extra matter, with masses far of being
accessible by any experiment, could equally well change the
conventional logarithmic running and force unification of the
gauge couplings at the required scale. Of course, intermediate
scale unification could in principle trigger a number of
phenomenological problems, such as fast proton decay. Also, some
nice features of the Minimal Supersymmetric Standard
Model (MSSM) unification at $10^{16}$GeV, among them the radiative
electroweak  breaking, could be problematic in principle.

In this short note we would like to investigate the changes that
the MSSM beta functions should suffer in order to achieve gauge
coupling unification at $10^{10-13}$GeV.  We further determine the
extra matter fields which make gauge couplings merge at an
intermediate energy  and show that such spectra may appear in the
context of specific string unified models which can in principle
avoid fast proton decay.  We also examine the conditions in order
to achieve radiative breaking of the electroweak symmetry, keeping
of course the top mass in its experimental value.

In the context of heterotic superstring theory,  the value of the
string scale is determined by the relation  $M_{string}/M_P =
\sqrt{a_s/8}$ where $a_s$ is the string coupling. This relation
gives a value some two orders of magnitude above  unification
scale  predicted by the MSSM  gauge coupling  running.  On the
contrary,  in the case of type I models, for example,  this ratio
depends on the values of the dilaton field and the
compactification scale. Choosing appropriate values for both
parameters it may be possible to lower the string scale.

The gravitational and gauge kinetic terms of the Type-I superstring
action are
\ba
S&=& \int\frac{d^{10}x}{(2\pi)^7}\left(e^{-2\phi_I}
\frac{1}{{\alpha'}^4}{\cal R}
+e^{-\phi_I}\frac{1}{4\,{\alpha'}^3}{\cal F}^2+\cdots
\right)\nonumber
\ea
where $\alpha'=M_I^{-2}$, with the string scale now being denoted
by $M_I$, while $e^{\phi_I}\equiv\lambda_I$ is the dilaton
coupling and $\phi_I$ the dilaton field.

 Consider now 6 of the 10 dimensions compactified on a 3 two-torii
$T^2\times T^2 \times T^2$ with radii $R_1,R_2,R_3$. Then, the
compactification volume is $V= \Pi (2\pi R_i)^2$. Assuming the
simplest case with isotropic compactification $R_1=R_2= R_3\equiv
R$, with the compactification scale $M_C=1/R$, the 4--d effective
theory obtained from the above action is
\ba
S&=& \int\frac{d^{4}x}{(2\pi)^7}\left(e^{-2\phi_I}
\frac{(2\pi\,R)^6}{{\alpha'}^4}{\cal R}
+e^{-\phi_I}\frac{(2\pi\,R)^6}{4\,{\alpha'}^3}{\cal F}_{(9)}^2
+\cdots \right)\label{action}
\ea
In the above, ${\cal F}_{(9)}$ is the 9--brane field strength of
the gauge fields while dots include similar terms for $7, 5, 3$
branes. The gauge fields and the various massless states arising
from non--winding open strings live on the branes while graviton
lives in the bulk\cite{jp}. {}From the action\ref{action} one
obtains the following expressions. The gravitational constant
$G_N$ is related to the first term and is given by
\ba
\frac 1G_N\,\equiv\,M_P^2 &=&\frac{8(2\pi
R)^6}{(2\pi)^6\alpha'^4}e^{-2\phi_I}=
           \frac{8}{\lambda_I^2}\frac{M_I^8}{M_C^6}
\ea
The gauge coupling is extracted from the field strength  term
${\cal F}_{(9)}$ of the gauge fields in the 9--brane in (\ref{action})
and is given by
\ba
 \frac{4\pi}{g_9^2}\,\equiv\,\frac{1}{\alpha_9}&=&
\frac{4\pi (2\pi R)^6}{(2\pi)^7\alpha'^3}e^{-\phi_I}=
            \frac{2}{\lambda_I}\frac{M_I^6}{M_C^6}
\label{a9}
\ea
where $g_9$ is the 9--brane coupling constant.
Combining the above two equations, one also obtains the relation
\ba
G_N&=& \frac{\lambda_I}4 \frac{a_9}{M_I^2}=\frac{\lambda_I}4
\alpha_9\alpha'
\label{GN}
\ea
It can be checked that for a $p$--brane in general, the formula
(\ref{a9}) generalizes as follows\cite{li,imr}
\ba
\alpha_p&=&\frac{\lambda_I}2 \left(\frac{M_C}{M_I}\right)^{p-3}
\label{ap}
\ea
Then, the formula (\ref{GN}) for the gravitational coupling constant
becomes
\ba
G_N=\frac{1}{M_{P}^2}=\frac{\lambda_I}4\alpha_p\alpha'
     \left(\frac{M_C}{M_I}\right)^{9-p}
\label{GNp}
\ea
The string unification scale may be also given in terms of the
compactification scale and the $p$--brane coupling as follows
\ba
M_I&=&\frac{\alpha_p}{\sqrt{2}}\left(\frac{M_C}{M_I}\right)^{6-p}M_P
\label{stsc}
\ea
{}From the last three expressions, it is clear that the
compactification scale $M_C\sim 1/R$ is rather crucial for the
determination of the string scale. We may explore the various
possibilities by solving for $M_I$ and $\lambda_I$ in terms of the
compactification scale and obtain the following relations
\ba
M_I&=&
\left(\frac{\alpha_p}{\sqrt{2}}M_C^{6-p}M_P\right)^{1/(7-p)}
\label{mip}\\ \lambda_I&=&2\alpha_p\left(\frac{\alpha_p}{\sqrt{2}}
\frac{M_P}{M_C}\right)^{\frac{p-3}{7-p}} \label{lip}
\ea
In terms of $\lambda_I$, the string scale for any $p$--brane  is
also written as follows
\ba
M_I&=&\left(\frac{\lambda_I}{2\sqrt{2}} M_P M_C^3\right)^{\frac 14}
\ea
where all the $p$--dependence is absorbed in $\lambda_I$. In order
to remain in the perturbative regime, we should impose the
condition $\lambda_I\le {\cal O}(1)$. From the last expression it
would seem natural to assume $M_I\sim M_C$ and demand that
$\lambda_I\ll 1$ to obtain a small string scale. However, this is
not a realistic case since from relation (\ref{ap}) we would also
have $\alpha_p\ll 1$, i.e., an extremely low initial value for the
gauge coupling. {}From (\ref{lip}) it can be seen that  viable
cases arise either for $p\le 3$ or $p > 7$.

In what follows, we  wish to elaborate further the case where the
string scale lies in the intermediate energies define by the
geometric mean $\sqrt{M_P M_W}\sim 10^{11}$GeV. The weak coupling
constraint on $\lambda_I$ above suggests that the effective field
theory gauge symmetry is more naturally embedded in a 3-- or
9--brane. Taking into consideration these remarks, the
corresponding compactification scale can be extracted from the
above formulae. In Table 1 we give some characteristic values of
the $M_I, M_C$ and $\lambda_I$ for the 2--, 3-- and 9--brane case
We assume that $\alpha_p\sim 1/20$ which, as we will see in the
next section, is indeed the correct value of the  unified coupling
for a unification scale around $M_U\sim 10^{10-13}$GeV.
%%%%   %%%%%
\begin{equation}
\begin{array}{|c|c|c|c|}
\multicolumn{4}{c}{Table\quad 1}\\
\hline
  p&\lambda_I& \log_{10}(M_I)&\log_{10}(M_C)\\ \hline\hline
  %2&    0.020& 13.7     & 13 \\ \hline
  %2&    0.012& 12.9     & 12 \\ \hline
  %2&    0.008& 12.1     & 11 \\ \hline
  3&    1/10 & 13.9     & 13 \\ \hline
  3&    1/10 & 13.1     & 12 \\ \hline
  3&    1/10 & 11.4     & 11 \\ \hline
  3&    1/10 & 11.6     & 10 \\ \hline
  9&    \ll 1& 12.8     & 12 \\ \hline
  9&    \ll 1& 11.2     & 11 \\ \hline
  9&    \ll 1&  6.7     & 10 \\ \hline
\end{array}
\end{equation}
{}From the above Table, it is clear that the requirement to remain
in the perturbative regime is satisfied in all cases considered
above. However, for the case of 9--branes, the dilaton coupling is
extremely small. On the contrary, the in the case of 3--branes
this coupling takes reasonable values, in fact its value is fixed
through the relation (\ref{ap}), $\lambda_I = 2\alpha_3$, being
independent of the ratio $M_C/M_I$. Therefore, the embedding of
the gauge group in the 3--brane looks more natural\cite{li,imr}.

%%%%%   RGEs %%%

\section{Renormalization Group Analysis}
In this section we will explore the possibility of modifying
the MSSM $\beta$--functions in order to implement the intermediate
scale  unification scenario. Next, we will give examples of
matter multiplets which fulfill the necessary conditions.     For
simplicity, we will assume in the following that the compactification
scale is the same as the string scale.
We begin by writing down the (one-loop) running of the gauge
couplings
\begin{equation}
\frac{1}{\alpha_i(M)}=\frac{1}{\alpha_U}+
\frac{\beta_i}{2\pi}\log\frac{M_U}{M_{SB}}+
\frac{\beta_i^{NS}}{2\pi}\log\frac{M_{SB}}{M},\quad\quad i=1,2,3
\label{rge}
\end{equation}
where $M_U$ is the unification scale and $M_{SB}$ is the SUSY
breaking scale and we have of course $M_U>M_{SB}>M$. In the
equation above, we have assumed that
\begin{itemize}
\item
the three gauge couplings unify at $M_U$ ($\alpha_i(M_U)=\alpha_U$)
\item
extra matter, possibly remnants of a GUT, appears in the region
between $M_U$ and $M_{SB}$: $\beta_i=\beta_i^S+\delta\beta_i$,
where $\beta_i^S=(33/5,1,-3)$ is the MSSM $\beta$ functions, and
\item
in the region between $M_{SB}$ and $M_Z$ we have the (non-SUSY) SM
(although with two higgs instead of one) and the corresponding
$\beta_i^{NS}=(4.2,-3,-7)$ functions.
\end{itemize}

By choosing $M\equiv M_Z$ in (\ref{rge}), we can solve the system
of these three equations with respect to $(M_U,M_{SB},\alpha_U)$
as functions of  the $\delta\beta_i$'s, taking the values of
$\alpha_i(M_Z)$ from experiment. In this sense, the
$\delta\beta_i$'s are treated as free continuous parameters.
However, when a specific GUT is chosen, these free parameters take
discrete values depending on the matter content of the GUT
surviving under the scale $M_U$. Solving therefore (\ref{rge}) we
get
\begin{eqnarray}
t_{SB}&=&\frac{
  \delta\beta_{jk}(2\pi\delta(\alpha^{-1})_{ji}+\delta\beta_{ji}^{NS}t_Z)-
  \delta\beta_{ji}(2\pi\delta(\alpha^{-1})_{jk}+\delta\beta_{jk}^{NS}t_Z)}
{-\delta\beta_{ji}(\delta\beta_{jk}^{NS}-\delta\beta_{jk})+
  \delta\beta_{jk}(\delta\beta_{ji}^{NS}-\delta\beta_{ji})}
\nonumber\\
\nonumber \\ 
t_U&=&\frac{2\pi\delta(\alpha^{-1})_{ji}+\delta\beta_{ji}^{NS}t_Z-
                       (\delta\beta_{ji}^{NS}-\delta\beta_{ji})t_{SB}}
         {\delta\beta_{ji}}
\nonumber\\
\frac{1}{\alpha_U}&=&\frac{1}{\alpha_i}-\frac{\beta_i^{NS}}{2\pi}(t_{SB}-t_Z)-
                            \frac{\beta_i}{2\pi}(t_U-t_{SB})
\nonumber
\label{solution}
\end{eqnarray}
where $t_{SB,U,Z}$ is the logarithm of the corresponding scales,
$\delta p_{ij} = p_j-p_i$ and $i,j,k$ should be different.
Although we have not written explicitly
the unknowns w.r.t. the $\delta\beta_i$'s (only $t_{SB}$ is given
explicitly) it is obvious that $t_U$, $t_{SB}$ and $\alpha_U$
depend only on the differences of $\beta_i$'s. Therefore, if a
certain solution $(t_U,t_{SB},\alpha_U)$ is obtained by using
specific values for $(\delta\beta_1, \delta\beta_2
,\delta\beta_3)$, the same solution is obtained for
$(\delta\beta_1+c, \delta\beta_2+c,\delta\beta_3+c)$ where $c$ is
an arbitrary constant.

By putting the following constraints
\begin{equation}
10^{10}<M_U/GeV<10^{13},\quad\quad 10^3<M_{SB}/GeV<3\cdot 10^3
\label{constraints}
\end{equation}
we plot in Fig.1 the acceptable values of
$(\delta\beta_1,\delta\beta_2)$ for $\delta\beta_3=0$. The four
``lines'' correspond to the four combinations
\[
(\alpha_3(M_Z),s^2_W\theta(M_Z))=(0.11,0.233),(0.11,0.236),(0.12,0.233),
(0.12,0.236)
\]
Translating the ``lines'' by an amount $c$ in both directions,
the corresponding figure for $\delta\beta_3=c$ appears.

%%%%%%%%%%%%%
\begin{figure}
\begin{center}
\psfig{figure=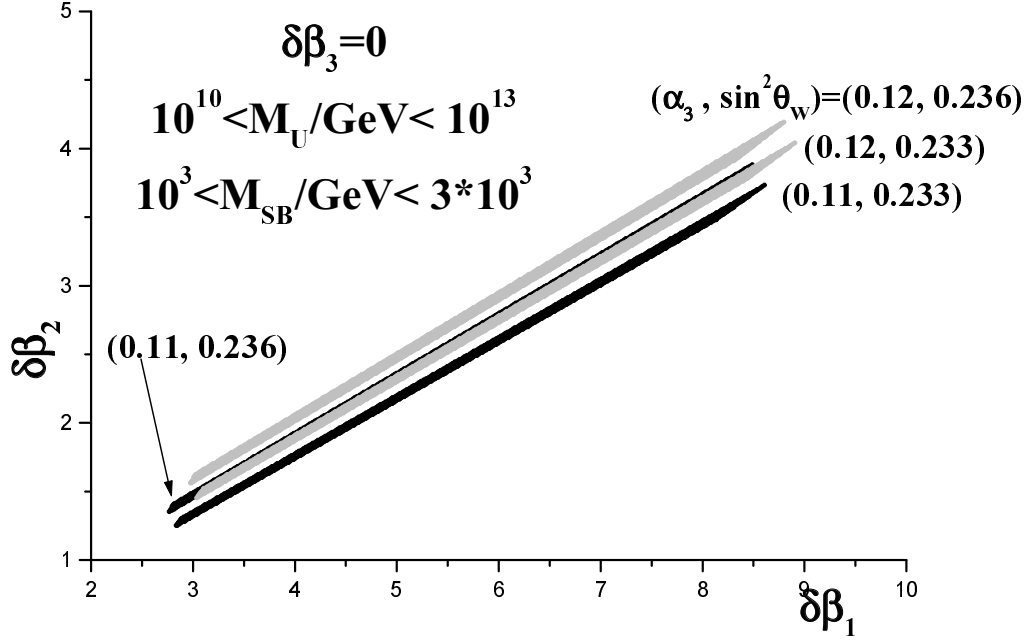,width=5cm}
\end{center}
\caption{The allowed region of the ($\delta\beta_1,\delta\beta_2$)
space, for $\delta\beta_3=0$, in order to achieve unification
in the region $10^{10}-10^{13}$ GeV, while the supersymmetry breaking
is in the region $1-3$ TeV. Four different ($\alpha_3,sin^2\theta_W$)
pairs are shown.}
\end{figure}
%%%%%%%%%%%%

In Fig.2 we plot the inverse of the unification coupling,
$\alpha_U^{-1}$, versus $(\delta\beta_1,\delta\beta_2)$, for
$\delta\beta_3=0$. Again, since $\alpha_U$ is one of the three
unknowns of (\ref{rge}), we can easily have the required
$\alpha_U$ for any value of $\delta\beta_3$. We see therefore a
slight increase of the unification coupling with respect to the
MSSM one ($\sim 1/24$). As far as the unification scale $M_U$ and
the SUSY breaking scale $M_{SB}$ are concerned, there is a
tendency to decrease as $\delta\beta_1$ gets bigger, while the
opposite happens for $\delta\beta_2$.

%%%%%%%%%%%%%
\begin{figure}
\begin{center}
\psfig{figure=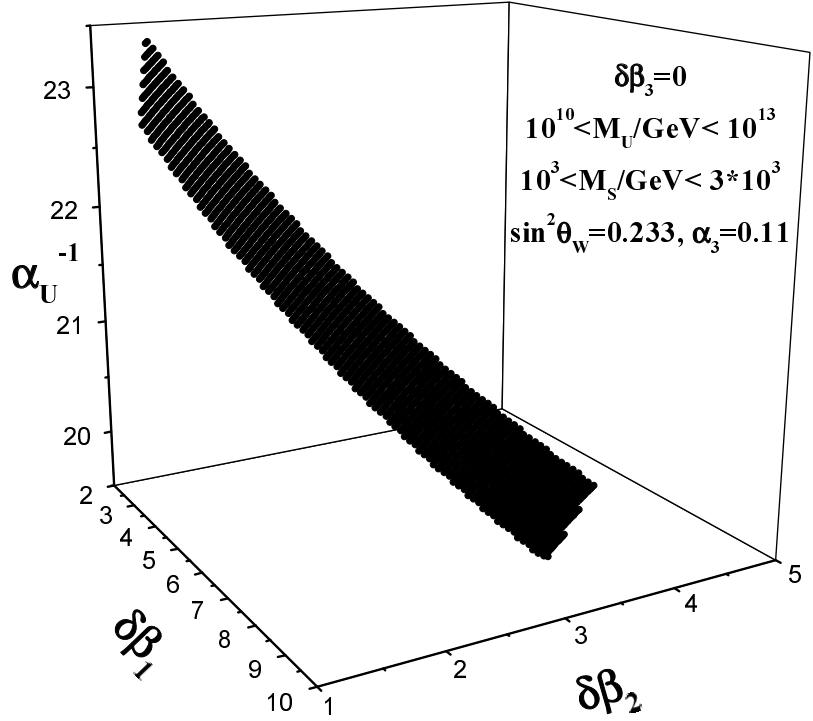,width=10cm}
\end{center}
\caption{The inverse of the unified gauge coupling as a function
of $(\delta\beta_1,\delta\beta_2)$, for $\delta\beta_3=0$ and the same
constraints from $M_U$ and $M_{SB}$ as in Fig.1.}
\end{figure}
%%%%%%%%%%%%

Let us try to find the acceptable values for a specific GUT model,
namely the $SU(4)\times SU(2)_L\times SU(2)_R$. In this case, we
assume that the breaking to the standard model occurs directly at
the string scale $M_I=M_U$, so that the gauge couplings $g_L, g_R,
g_4$ attain a common value $g_U$.  The massless  spectrum of the
string model -- in addition to the three families and the standard
higgs fields -- decomposes to the following
$SU(3)\times SU(2)_L\times U(1)_Y$ representations\cite{alt}
\begin{eqnarray}
n_2 & \rightarrow & (1,2,\pm 1/2),\quad\quad
                   n'  \rightarrow (1,1,\pm 1/2)\nonumber\\
n_3 & \rightarrow & (3,1,\pm 1/3),\quad\quad
                   n_3'\rightarrow  (3,1,\pm 1/6)\nonumber\\
n_{31} & \rightarrow & (3,1,\pm 2/3),\quad\quad
                   n_L  \rightarrow  (1,2,0)\nonumber
\end{eqnarray}
In the above, $n_{2},\, n_{3},\cdots, $ represent the number of
each multiplet which appears in the corresponding parenthesis with
the quantum numbers under $SU(3)\times SU(2)_L\times U(1)_Y$. In
this case, the $\delta\beta_i$'s are given explicitly
\begin{eqnarray}
\delta\beta_1&=&\frac{n_2}{4}+\frac{n'}{4}+\frac{n_3}{3}
+\frac{n_3'}{12}+\frac43
n_{31} \nonumber\\ \delta\beta_2&=&\frac{n_2+n_L}{2} \nonumber\\
\delta\beta_3&=&\frac{n_3+n_3'+n_{31}}{2} \label{dbis}
\end{eqnarray}
In the specific GUT model the above $n$'s are even
integers. Therefore, we see that $\delta\beta_{2,3}$ are integers
while $\delta\beta_1$ can change by steps of $1/6$. In that case
only the following 3 points are acceptable in all the region
allowed by the constraints on $sin^2\theta_W(M_Z)$ and
$\alpha_3(M_Z)$ having put earlier (keeping $\delta\beta_3=0$)
\[
(\delta\beta_1,\delta\beta_2,\delta\beta_3)= (4,2,0),\quad
(6.5,3,0), \quad (8.5,4,0)
\]
 Several possible sets of $n$'s
can generate the above changes in the $\beta$--functions.
Again, as was mentioned above, acceptable values for higher
$\delta\beta_3$ can be obtained in a straightforward manner
\[
(4+c,2+c,c),\quad (6.5+c,3+c,c), \quad (8.5+c,4+c,c)
\]
where $c$ is an integer but not any integer, since eq(12) should
be satisfied for even $n's$. It is easy to see from these
equations that we need to change $\delta\beta_3$ by 3 units to
find an acceptable solutions for the $n's$
\begin{equation}
\begin {array}{cccc}
(4,2,0) & (7,5,3) & (10,8,6) & ...\\
(6.5,3,0) & (9.5,6,3) & (12.5, 9 ,6) & ...\\
(8.5,4,0) & (11.5,7,3) & (14.5, 10,6) & ...
\end{array}
\end{equation}
 Of course, to these values correspond
different sets of $n$'s and obviously as the $\delta\beta_i$'s
increase more and more possible sets appear. We give the possible
$n's$ for the three acceptable cases with $\delta\beta_3=0$
\begin{equation}
\begin{array}{||l||r|r|r||r|r|r|r||r|r|r|r|r||}
\multicolumn{13}{c}{Table\quad 2}\\
\hline
     &\multicolumn{3}{c||}{(4,2,0)}&
                   \multicolumn{4}{c||}{(6.5,3,0)}&
                                  \multicolumn{5}{c||}{(8.5,4,0)}\\
\hline\hline n_2  & 0 & 2 & 4 & 0 & 2 & 4 & 6 & 0 & 2 & 4 & 6 &
8\\ \hline n_L  & 4 & 2 & 0 & 6 & 4 & 2 & 0 & 8 & 6 & 4 & 2 & 0\\
\hline n'   &16 &14 &12 &26 &24 &22 &20 &34 &32 &30 &28 &26\\
\hline
\end{array}
\end{equation}
while of course $n_3=n_{31}=n'_3=0$

We have also checked whether the radiative breaking of the electroweak
symmetry is still applicable. In other words, we have used the
coupled differential equation
 governing the
running of the mass squared parameters of the scalars and checked
that only $\tilde m^2_{H_2}$ becomes negative at a certain scale.
This scale depends of course on the chosen $\delta\beta _i$'s, but
stays in the region between $10^5-10^7$GeV.

In conclusion, we have checked the possibility of intermediate
scale ($10^{10-13}$ GeV) gauge coupling unification, using the
traditional logarithmic running, i.e. without incorporating the
power-law dependence on the scale coming from the Kaluza-Klein
tower of states. We have showed that this kind of unification can
be achieved with small changes of the $\beta$-functions of the
MSSM gauge couplings, which can be attributed to matter remnants
of superstring models. We have applied the above to the successful
$SU(4)\times SU(2)_L\times SU(2)_R$ model (which is safe against
proton decay even in this intermediate scale), and found the
necessary extra massless matter and higgs fields needed. Finally
we have checked that the radiative electroweak breaking of the
MSSM still persists, driving the mass squared of the higgs to
negative values at the scale $\sim 10^{5-7}$ GeV, while all others
scalar mass squared parameters stay positive.

\end{document}